\documentclass[%
 aip,
 amsmath,amssymb,
 reprint,%
]{revtex4-1}

\usepackage{graphicx}
\usepackage{dcolumn}
\usepackage{bm}

\usepackage[utf8]{inputenc}
\usepackage[T1]{fontenc}
\usepackage{mathptmx}
\usepackage{etoolbox}
\usepackage{xcolor}
\newcommand{\nc}{\newcommand}
\nc{\nn}{\nonumber}
\nc{\txt}{\textrm}
\nc{\txtsup}{\textsuperscript}
\nc{\txtsub}{\textsubscript}
\nc{\calL}{\mathcal{L}}
\nc{\U}{\mathcal{U}}
\nc{\T}{\mathcal{T}}
\nc{\E}{\mathcal{E}}
\nc{\calH}{\mathcal{H}}
\nc{\SARKAR}[1]{\textcolor{black}{#1}}
\nc{\YD}[1]{\textcolor{black}{#1}}
\newcommand{\orcid}[1]{\href{https://orcid.org/#1}{\includegraphics[width=8pt]{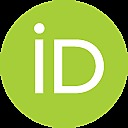}}}

\makeatletter
\def\@email#1#2{%
 \endgroup
 \patchcmd{\titleblock@produce}
  {\frontmatter@RRAPformat}
  {\frontmatter@RRAPformat{\produce@RRAP{*#1\href{mailto:#2}{#2}}}\frontmatter@RRAPformat}
  {}{}
}%
\makeatother
\begin{document}


\title{Temperature-dependence of the chirality-induced spin selectivity effect - experiments and theory}
\author{Seif Alwan}
\affiliation{Department of Chemistry, Ben-Gurion University of the Negev, Beer Sheva, 84105,  Israel
}

\author{Subhajit Sarkar \orcid{0000-0002-7260-5100}}
\email{subhajit@post.bgu.ac.il}
\affiliation{Department of Chemistry, Ben-Gurion University of the Negev, Beer Sheva, 84105,  Israel
}
 \altaffiliation{School of Electrical and Computer Engineering, Ben-Gurion University of the Negev, Beer Sheva, 84105,  Israel}

\author{Amos Sharoni \orcid{0000-0003-3659-0818}}
\email{amos.sharoni@biu.ac.il}
\affiliation{Department of Physics, Institute of Nanotechnology \& Advanced Materials, Bar-Ilan University, Ramat-Gan, 5290002, Israel
}

\author{Yonatan Dubi \orcid{0000-0002-8988-4935}}
\email{jdubi@bgu.ac.il}
\affiliation{Department of Chemistry, Ben-Gurion University of the Negev, Beer Sheva, 84105,  Israel
}
\affiliation{Ilse Katz Center for Nanoscale Science and Technology, Ben-Gurion University of the Negev, Beer Sheva, 84106,  Israel
}

\date{\today}

\begin{abstract}
The temperature dependence of the chirality-induced spin-selectivity (CISS)  effect can be used to discriminate between different theoretical proposals for the mechanism of the CISS effect. Here we briefly review key experimental results and discuss the effect of temperature in different models for the CISS effect.  We then focus on the recently suggested spinterface mechanism and describe the different possible effects temperature can have within this model. Finally, we analyze in detail recent experimental results from Qian et al. [Nature  606, 902–908 (2022)] and demonstrate that, opposite to the original interpretation by the authors, these data actually indicate that the CISS effect increases with decreasing temperature. Finally, we show how the spinterface model can accurately reproduce these experimental results. 
\end{abstract}

\maketitle
\section{Introduction}
As electrons pass through chiral molecules, the resulting current (either generated optically or electrically) can become spin-polarized \cite{Naaman12,Naaman15,naaman2019chiral,naaman2020chiral}, implying that in effect spins of different species have different transport properties when interacting with a chiral moiety. This leads to the so-called chirality-induced spin selectivity (CISS) effects. The CISS effect has been observed in numerous experimental platforms and settings, and can be used for various spintronic applications. \cite{Banerjee-Ghosh18,ghosh2019controlling,kumar2017magnetless,dor2017magnetization,al2018single,torres2020reinforced,abendroth2019spin,ko2022twisted}. The observation of the CISS effect in molecular junctions is obtained by measuring the I-V curves through a chiral molecular moiety when the junction is comprised of one metallic (typically Au) electrode and one ferromagnetic (typically Ni) electrode. The hallmark of the CISS effect in molecular junctions is a different I-V response of the system when the magnetization of the Ni electrode is reversed, from being parallel to being anti-parallel to the direction of the current (typically the molecular axis \cite{clever2022benchmarking}). 
 
 Surprisingly, despite the fact that over a decade has passed since its discovery, the question of the physical origin of the CISS effect remains open \cite{Evers20,naaman2020chiral}. Various theoretical studies proposed that the CISS effect originates from spin-orbit coupling (SOC) inside the chiral molecule \cite{Guo12a,Gutierrez12,Guo12b,Gutierrez13,Guo14a}. However (as was recently pointed out explicitly \cite{Ghazaryan20,naaman2019chiral,Evers20}), these explanations require a huge  (several orders of magnitude) renormalization of the molecular SOC, and while electron correlations \cite{Fransson19,li2020chiral} or vibrations \cite{fransson2020vibrational,fransson2021charge,Li2020,Du2020,Zhang2020} may reduce this normalization to some extent, realistic values for the SOC seem to predict a much smaller CISS effect than experimentally observed. While other theoretical ideas are beginning to emerge  \cite{dalum2019theory,Vittmann2022,teh2022spin},  
To date, none of the suggested theories is able to quantitatively reproduce the  experimental data.
 
Recently, an alternative theoretical explanation to the CISS effect was suggested \cite{alwan2021spinterface}, based on the so-called ``spinterface" mechanism. As arises from this approach, the CISS effect is a result of interplay between the (spin-orbit induced) surface magnetization in the metal electrode and spin-imbalance in the molecule, which interact via a spin-torque. Due to  spin-transfer torque, the surface magnetization obtains a preferred direction (parallel or anti-parallel to the molecular axis), determined by the molecular chirality, thus forming a spin-dependent barrier for electrons entering the molecule. This theoretical approach can be used to reproduce the experimental data (the magnetization-dependent I-V currents) with remarkable accuracy \cite{dubi2022spinterface,Yang2023}. 


\section{Temperature dependence of the CISS effect - a brief review }
The emergence of various theoretical suggestions, all plausible to the CISS effect, requires various experimental means to distinguish between them. One possible ``experimental knob" that can help differentiate between different theories for the CISS effect is the temperature. Changing the temperature while observing the CISS effect seems only natural \cite{tsymbal2003spin,Miyazaki98}, yet there are only a few precious experimental results in this direction. Perhaps the earliest example is in Ref.~\onlinecite{kumar2013device}, where it was demonstrated that spin-polarization (obtained from a voltage difference between a magnetic layer and a metallic layer on which chiral molecules were placed) decreases with increasing temperatures. 

More recently, several works reported a temperature dependence of the CISS effect, yet they disagree. Yang et al. \cite{Yang2023} demonstrated a clear decrease of the CISS effect with increasing temperature, which was explained (and fitted quantitatively) with the so-called spinterface mechanism for te CISS effect. On the other hand, Das et al. \cite{das2022temperature} showed an increase of the effect at high temperatures, explained via the so-called phonon mechanism \cite{fransson2020vibrational} (yet not fitted quantitatively, see discussion in \cite{dubi2022spinterface}). Kondou et al. \cite{kondou2022chirality} also show an increase in the signal when temperatures increase, via measurements of the current-in-plane magnetoresistance, however in this experiment current does not pass through the molecular layer. Finally, Qian et al. \cite{qian2022chiral} claimed to show a decreasing CISS effect when temperature is increased, via measurements of the CISS effect in chiral molecular intercalation superlattices. However, as we show in detail in Sec.~IV, the data of Qian et al. actually points to the reverse, namely that in their system the CISS effect is in fact largest at low temperatures and decreases as the temperature is increased, until the CISS signal vanishes at high enough temperature. 

From the theoretical perspective, one can divide the proposed mechanisms for the CISS effect into two categories. The first is mechanisms that depend on interactions between vibrations (or phonons) and electrons in chiral molecules as a means to enhance the (typically weak) spin-orbit coupling \cite{fransson2020vibrational,das2022temperature, Li2020, Du2020, Zhang2020}. Since the excitation of vibrations increases with temperature, it follows that as temperature increases, more vibrational modes are excited, leading to an enhanced CISS effect. All the mechanisms that do not depend on vibrations belong to the second category. Naturally, if the vibrations are not involved in the mechanism leading to the CISS effect, one would expect that thermal fluctuations would destabilize it by competing with the interactions that give rise to it. 

In the next sections, we describe in detail the temperature-dependence of the CISS effect, as arises from the spinterface mechanism \cite{alwan2021spinterface,dubi2022spinterface}. We then discuss the work of Qian et al. \cite{qian2022chiral}, and show that, (i) the original interpretation of the data contains some possible flaws, which, when corrected, actually indicate that their data shows a reduction of the CISS effect with increasing temperature, and (ii) the spinterface mechanism can quantitatively reproduce the experimental results of Qian et al. remarkably well. 

\section{Temperature-dependence from the spinterface model} \label{sec:spinterface_model}

Within the spinterface model \cite{alwan2021spinterface,dubi2022spinterface}, the CISS effect arises from the stabilization of surface magnetization at the interface between the metallic electrode and the chiral molecule. The stabilization of the surface magnetization comes from spin-torque (or possibly spin-exchange) interactions between the metallic orbitals and the molecular states, and the symmetry is broken due to the (albeit small) current-induced solenoid magnetic field. Finally, the effect of the surface magnetization on transport comes from the fact that in the metallic electrode there is substantial spin-orbit interactions, leading (in a mean-field description) to a spin-dependent effective chemical potential, or equivalently to a spin-dependent tunneling barrier. 

In order to evaluate the temperature dependence of the CISS effect, it is useful to briefly describe the formulation of the spinterface mechanism (described in full in Ref.~\onlinecite{alwan2021spinterface}). If the I-V curves through the junction when the ferromagnetic (e.g. Ni) electrode is not magnetized is given by $I(V)$, then the magnetization-dependent currents $I_s (V)$ (where $s=\pm 1$ stand for Ni magnetization parallel or anti-parallel to the chiral molecular axis) are given by 
\begin{equation}
I_s(V)=I(V+s \alpha_A \cos(\theta_M))~~,
\label{Eq1-spinterface} \end{equation}
where $\alpha_A$ is the spin-orbit interaction in the metallic electrode and $\cos(\theta_M)$ is the average tilt angle of the surface magnetization from the molecular axis. 

The average tilt angle $\cos(\theta_M)$ is evaluated self-consistently via 
\begin{equation}\label{theory_Eq.1}
 \cos(\theta_M)=\mathcal{B}\left[\frac{\mu B_{eff}}{k_B T}\right]~.
\end{equation}
 Here $\mathcal{B}[x]=J_1(x)/J_0(x)$, where $J_n(x)$ are the Bessel functions of the first kind, $k_B$ is the Boltzmann constant and $T$ the temperature. The effective field is defined in terms of the total current $I$ through the junction, and the induced spin-density in the molecule $\Delta s=n_\uparrow-n_\downarrow$, through 
 \begin{equation}\label{theory_Eq.2}
 \mu B_{eff}= \alpha_0 I+\alpha_1 \Delta s ~,
 \end{equation}
where $I$ is the total charge current and $\alpha_0$ is a constant relating the current through the helical system (i.e. a tiny solenoid) to the generated field. Typically, $\alpha_0$ is very small \cite{DiVentra11}  and thus by itself this term cannot generate the CISS effect. However, the effect is amplified by the spin-torque interaction, manifested through the second term, where $\alpha_1$ is the interaction strength and $\Delta s$ is the spin-density in the molecule. Note that this formulation naturally couples the "magnetic polarizability" (namely the spin density generated in the molecule when it is placed in an electric field, or equivalently when current is passed through it) to the CISS effect, a connection which was demonstrated experimentally via the connection between the CISS effect and circular dichroism  \cite{Kulkarni2020highly}. The self-consistent nature of the above formulation comes about from the fact that both the total current $I$ and the spin-density $\Delta s$ depend on the spin-dependent shift in the chemical potential, $V\rightarrow V\pm \alpha_A \cos(\theta_M)$. 

This formulation shows two temperature dependencies. The first is a direct dependency which comes from Eq. ~\eqref{theory_Eq.2}, where the temperature appears in the denominator of the expression. This can be easily understood by noting that this formulation comes from averaging over a thermally fluctuating magnetic moments \cite{alwan2021spinterface}. Thus, it is natural that as the temperature increases, it becomes harder to ``stabilize" the surface magnetization. Mathematically, as $T$ increases the Bessel functions in Eq. ~\eqref{theory_Eq.2} decrease, leading to a diminishing $\cos(\theta_M)$ and a reduction of the CISS effect.   

The second, more subtle dependence, comes from the dependence of $I$ and, more importantly, $\Delta s$ on temperature. These dependencies are less universal, and may depend on the details of the specific system which shows the CISS effect. However, one can largely assume that $\Delta s$ decreases as temperature increases. This is indeed the case for the microscopic models examined in Ref.~\onlinecite{alwan2021spinterface}. 

\begin{figure}
    \centering
    \includegraphics[keepaspectratio=true,scale=0.1]{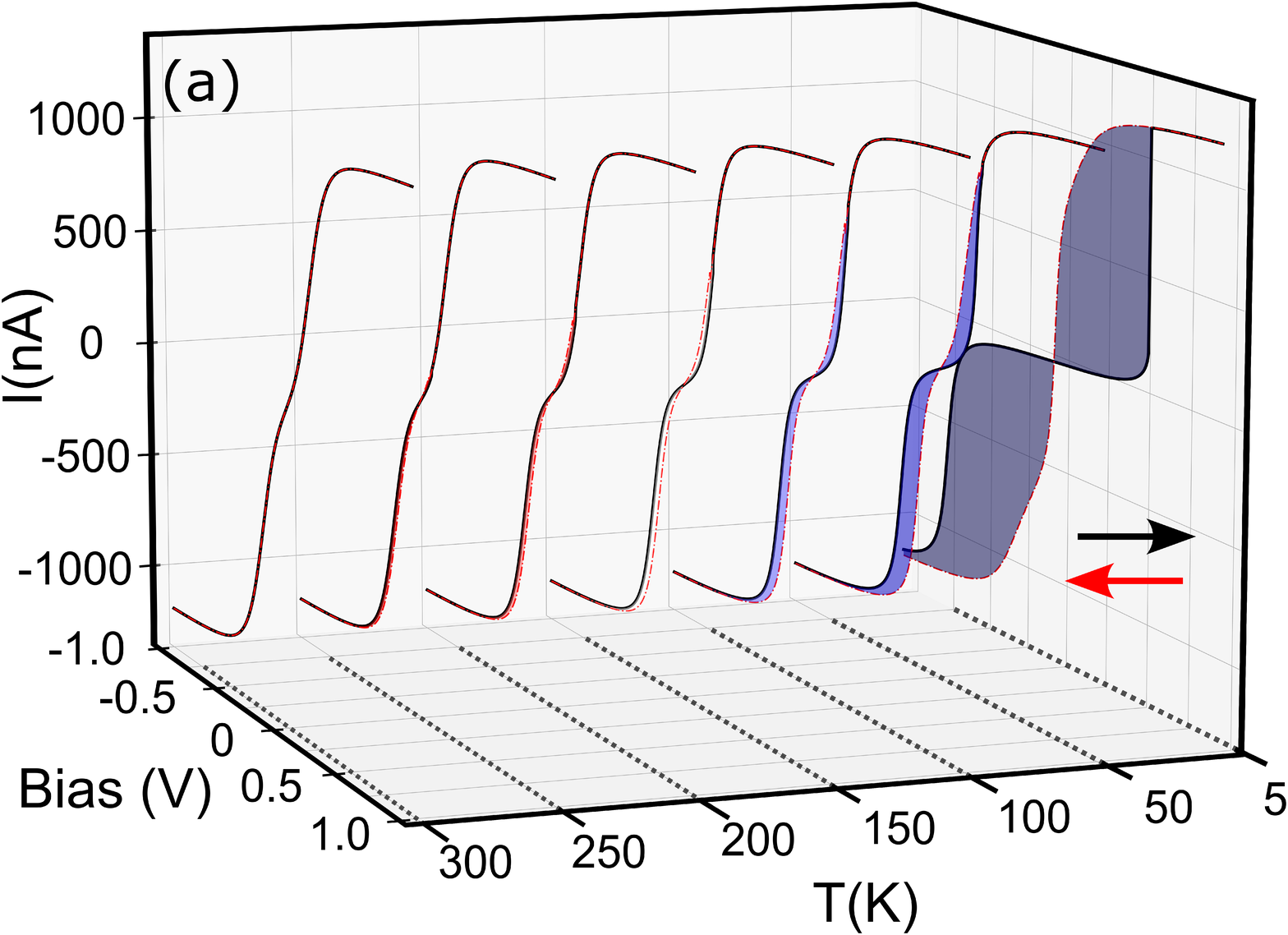}
    \includegraphics[keepaspectratio=true,scale=0.1]{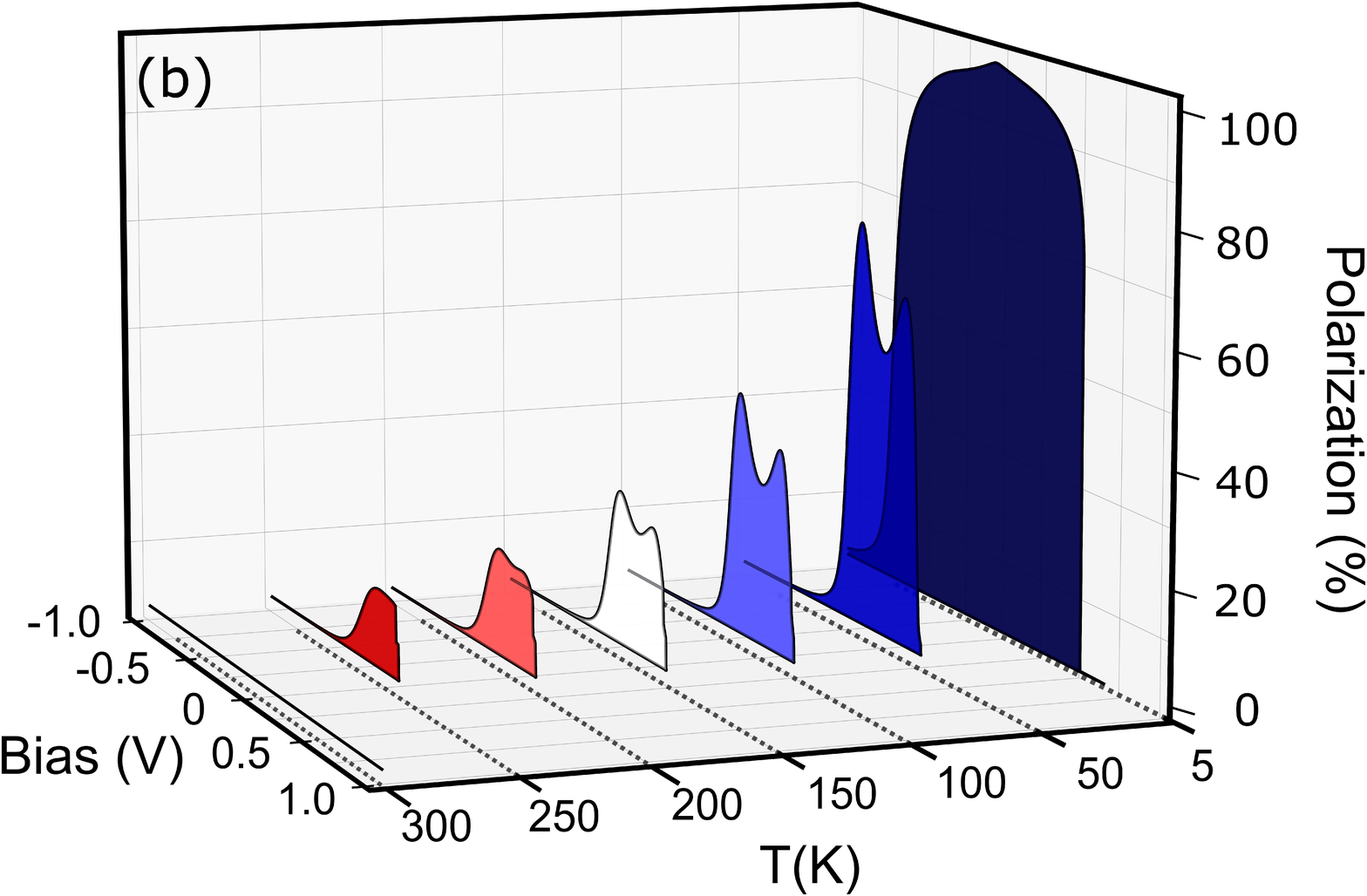}
    \caption{The temperature dependence of the CISS effect from the spinterface mechanism: (a) current-voltage curves for different Ni electrode magnetizations (solid lines - magnetization parallel to the molecular axis, dashed lines - magnetization antiparallel to the molecular axis) for different temperatures, ranging from 5K up to 300K. Filled area marks the differences between the currents, and vanishes with increasing temperatures, marking the diminishing of the CISS effect. (b) CISS polarization vs voltage corresponding to the currents shown in (a), demonstrating the decrease of the CISS effect. }
    \label{Fig:J_V_T}
\end{figure}

As a demonstration, in Fig.~\ref{Fig:J_V_T} we show how the CISS effect vanishes with increasing temperature, by plotting the current-voltage curves for different Ni electrode magnetizations (solid lines - magnetization parallel to the molecular axis, dashed lines - magnetization antiparallel to the molecular chiral axis) for different temperatures, ranging from 5K up to 300K. The system considered here is a single molecular orbital, and the currents are evaluated within the full temperature-dependent Landauer approach \cite{Book:Cuevas_Scheer10}. 
 
 We set the resonant energy at $\varepsilon=-0.1$ eV and level broadening at $\Gamma=0.01 eV$, and the CISS parameters are $\alpha_A=0.5 eV$, $\alpha_0=10^{-7}$ eV/nA and $\alpha_1=5\times 10^{-4}$ eV. Note that $\alpha_1$ here is very small (in realistic systems it is probably closer to $10^{-2}$ eV, see Ref.~\onlinecite{dubi2022spinterface}), which was chosen in order to demonstrate the temperature dependence: the difference between the magnetization currents (which is color-filled for clarity), which is the hallmark of the CISS effect, vanishes as the temperature increases.



Such behavior has been demonstrated experimentally very recently \cite{Yang2023} in molecular junctions with small molecules. The observation of the CISS effect at such small temperatures (where phonons are unimportant) corroborates the spinterface mechanisms for these systems. However, it is possible that for larger molecules (for example, biological molecules), vibrations play a dominant role in stabilizing the CISS effect. 

Before discussing the relevant experimental results in the next section, here we comment that while observing the temperature-dependence of the CISS effect from the magnetization-dependent currents is straightforward, {\sl quantifying}  it may be more complicated. The reason is that quantifying the CISS effect may actually depend on the theoretical modeling of the full system, including the properties of the ferromagnet electrodes, interfaces and non-magnetic electrodes. Hence, the quantification of the CISS temperature-dependence may depend on how the currents $I_s(V)$ (and possibly other parameters) depend on temperature. Put simply, if the CISS effect depends (via some mechanism) on a magnitude $X(V)$ (e.g. the currents), and $X(V)$ is temperature dependent, then the CISS effect will change with temperature even though the mechanism itself is temperature-independent, simply because $X$ depends on temperature. A similar point was made in Ref.~\onlinecite{dubi2022spinterface} regarding the length-dependence of the CISS effect. 

We thus reiterate a key message -- a theoretical description of the CISS effect in molecular junctions should be able to account for the basic raw data, namely the magnetization-dependent currents, and a qualitative (or even quantitative) agreement of any manipulated data must be rigorously justified. An example for such a case is described in the next section.  

\section{Temperature-dependence of the CISS effect from measurements in Chiral molecular intercalation super-lattices}
In this section we discuss in detail a recent report on the temperature-dependence of the CISS effect. In a recent paper Qian, et al., \cite{qian2022chiral} (which we dub Q22 hereafter for brevity) demonstrated chiral molecular intercalation superlattices (CMIS) as a new class of solid-state chiral material platform for exploring the CISS effect. 
Specifically, Q22 showed that CMIS can be used to accurately characterize the CISS effect and explore its dependence on temperature, and other material properties. They showed a very high degree of polarization of the spin-current $P(T)$, that monotonically decreased with increasing temperature, and the average conductance $G_{SI}(T)$ exhibited an Arrhenius behavior, viz., $G_{SI}(T) = G_0 e^{-\frac{\epsilon_{A}}{k_B T}}$  corresponding to a thermally activated hopping. The spin-polarized conductance $G_{S} (T)$, on the other hand, exhibited a non-monotonic dependence, increasing with temperature at low temperatures up to a certain temperature ($\sim 100$ K) and then decreasing.  

The authors of Q22 ascribed this non-monotonic behavior to two mechanisms (as stated in the description they provide below their Fig. 5): the increase up to $100$ K is due to electron-phonon interaction assisted increase of the spin-selectivity of the chiral molecule \cite{das2022temperature, fransson2020vibrational}, and subsequent decrease is due to the decrease in the electron spin-polarization of the ferromagnetic $\text{Cr}_{3}\text{Te}_4$ lead. Put simply, they attribute the rise in the polarization at low temperatures to the increase of the CISS effect (because of the phonon-mechanism for the CISS effect). Thus, the interpretation of Q22 supports the phonon-mechanism for the CISS effect.

In this section, based on the data of Q22, we show that the non-monotonicity of the spin-polarized conductance $G_{S} (T)$ may in fact arise from a different mechanism than that proposed by the authors. The low-temperature behavior comes from the Arrhenius envelope of the currents (and not the electron-phonon interaction assisted increase of the spin-selectivity of the chiral molecule), and the reduction at high-temperatures comes from the decrease of magnetism and polarization of the ferromagnetic electrode and from vanishing of the CISS effect at high temperatures. We show that this interpretation is consistent with the experimental findings reported. Based on this we argue that in the CMIS system, the CISS effect actually grows with decreasing temperature. 

We start by reiterating the definitions and central result pertaining to the temperature-dependence of the CISS effect of Q22. The authors measure their CMIS device with two different magnetization directions (parallel and anti-parallel to the direction of current flow), and obtain two currents, marked $I_{\text{high}}$ and $I_{\text{Low}}$, the difference between them representing the presence of a CISS effect.

The polarization is defined as $P =  \frac{(I_{\text{high}}-I_{\text{low}})}{(I_{\text{high}}+I_{\text{low}})} =I_S/I_{\text{tot}}$ where $I_{\text{tot}}$ is the average current (times $2$), and $I_S$ is the spin-polarized current. The spin-dependent (or CISS-) conductance is given by $G_S=I_S /V$. Their central result regarding the CISS effect is that  although $G_S \rightarrow 0$ as $T\rightarrow 0$, $P$ reaches a finite value ($60\%$). This apparent opposite behaviors of $P$ and $G_S$ with temperature has been attributed to an increase in the CISS with temperature increase at low temperatures, followed by a global decrease due to the tunneling effect in the material part.


\begin{figure}
    \centering
    \includegraphics[keepaspectratio=true,scale=0.45]{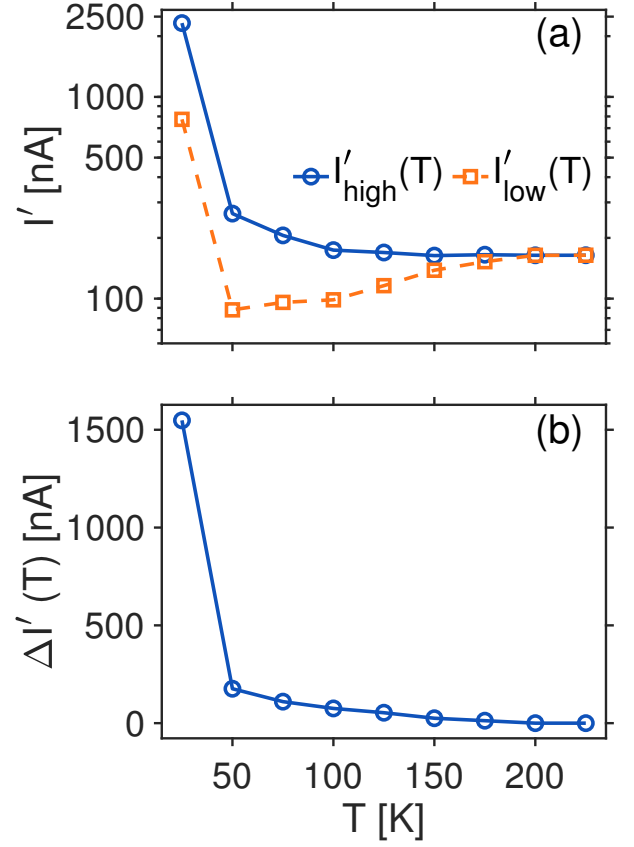}
    \caption{Current after filtering out the Arrhenius dependence: Currents after filtering out the Arrhenius dependence corresponding to the spin-up and down electrons $I^{\prime}_{\text{high}}$ (the blue circles connected by blue solid line) and $I^{\prime}_{\text{low}}$ (the orange squares connected by orange dashed line) in nA, respectively on a semi-log scale (a) and the difference between the Arrhenius filtered current $\Delta I^{\prime}$ in nA on a linear scale (b) as a function of temperature $T$ in K.}
    \label{Fig:1}
\end{figure}


Now, since $P$ and $G_S$ are proportional to one another (both having $I_S$ in the numerator), the only way this can happen is if $I_S$ and $I_{\text{tot}}$ decay to zero as $T\rightarrow 0 $ with the same functional form. According to Qian, et.al., \cite{qian2022chiral}, the average conductance $G_{\text{SI}} = \frac{I_{\text{high}} (T)+ I_{\text{low}} (T)}{2V}$ exhibits an Arrhenius form with $G_{\text{SI}} = G_0 e^{-\epsilon_A/k_B T}$ with an activation energy $\epsilon_A = 12 \text{meV}$, for a fixed (and of course temperature independent) applied bias $V$ (assuming $V=0.05$ eV, corresponding to 5 volt bias in Q22).

Therefore, the individual currents $I_{\text{high(low)}} (T)$ must also exhibit an Arrhenius component when $T\rightarrow 0$, i.e., $I_{\text{high(low)}}(T) = I^{\prime}_{\text{high(low)}} (T) e^{-\epsilon_A/k_B T}$. This is evident from Fig.~ext7(b-e) of Q22, and also supported by the fact that these currents are, after all, charge currents (see discussion in \cite{fransson2022chiral}), and therefore they should also be affected by the same mechanism which determines the Arrhenius behavior. Denoting $I^{\prime}_{\text{high(low)}} (T)$ as the Arrhenius normalized charge-current, the spin-conductance must be $G_{S} (T) = \left[ I^{\prime}_{\text{high}}(T) - I^{\prime}_{\text{low}} (T) \right] e^{-\epsilon_A/k_B T} = \Delta I^{\prime}(T) e^{-\epsilon_A/k_B T}$. This implies that the vanishing of $G_S$ with decreasing temperature comes from the Arrhenius part, and the part coming from the CISS effect is actually $\Delta I^{\prime}(T)$, the ``Arrhenius normalized" spin-polarized current. 

 To show this from the data, we digitally extract the temperature dependence of individual charge currents from Figs. 5a and ext7a of Q22. Fig. \ref{Fig:1}(a) shows the ``Arrhenius normalized" currents,  $I^{\prime}_{\text{high(low)}} (T)$, as a function of temperature $T$.  $I^{\prime}_{\text{high}}(T) $ decreases monotonically with increasing temperature, while $I^{\prime}_{\text{low}}(T) $ first decreases and then increases, indicating non-monotonicity. However, Fig. \ref{Fig:1}(b) shows that their difference $\Delta I^{\prime} (T)$ decreases monotonically with increasing temperature, indicating a monotonic decrease in the CISS effect. 
 
 Therefore, a consistent interpretation of the data (recalling that the polarization $P$ and the CISS conductance $G_S$ show distinctly different behaviors as $T\rightarrow 0$) is that the CISS effect is finite at low temperatures and decreases with increasing temperature,  while the total charge current increases with temperature, after being nearly zero at low temperatures. The competition between these two processes leads to non-monotonicity in $G_S$ but to a monotonic polarization $P$. 
 

The authors of Q22 interpreted their data using the Jullier model \cite{julliere1975tunneling}, which implies $G_S=G_T P_1 P_2$, where $P_1$ and $P_2$ are polarization of the Ferromagnetic lead and the equivalent polarization of the chiral molecular layers, respectively, and $G_T$ is the overall conductance envelope. Following the convention in their paper, the spin conductance can be directly expressed as $G_S=G_{SI} P$, where $P$ is related to the polarizations of the two layers and $G_T$ to $G_{SI}$.  The authors suggest, based on previous studies of different systems \cite{shang1998temperature} (Ref.~50 in  Q22) that $G_T$ is temperature-independent. However, this is in contradiction to their own data: in their system, $G_T$ is strongly temperature-dependent, and has an Arrhenius form which vanishes at low temperatures based on our picture. Therefore, their basic form of the Jullier's model, which assumes a temperature-independent conductance envelope is inconsistent with their data. The direct conclusion from this analysis and Fig.~\ref{Fig:1} is that the CISS effect in Q22 cannot be interpreted using the phonon-mechanism for the CISS effect.
 
However, the data provided by Q22 can be interpreted by a mechanism for the CISS effect in which the CISS effect is enhanced when the temperature is reduced, such as the aforementioned spinterface mechanism \cite{alwan2021spinterface}. To further corroborate this claim, we performed a calculation based on the spinterface theory of Ref.~\onlinecite{alwan2021spinterface}, modified to account for the specific Arrhenius-type I-V characteristics observed in the experiments performed by Q22. As we demonstrate, there is an excellent qualitative agreement between this theory and the experimental results. 

As described in Sec.~\ref{sec:spinterface_model}, for the spinterface theory one needs a model of currents and spin-densities. In the limit of weak coupling, compared to the temperature, between the molecular level and the electrodes, one can resort to the formulation of rate equations that simplifies the expressions for currents \cite{alwan2021spinterface}. For simplicity, we consider a single molecular level that mimics the complicated transport set-up fabricated in Q22. We treat the transport set-up corresponding to Q22 as a single level with an effective LUMO equal to the thermal activation energy. This way, we can separate the true thermal effects (i.e., the Arrhenius dependence found in Q22) from the spin-dependent effects.

The expression for magnetization-dependent currents is given by \cite{alwan2021spinterface}, 
\begin{eqnarray}\label{spin_dep_curr}
I_{\sigma} = \frac{\gamma_L \gamma_{R\sigma}}{\gamma_L + \gamma_{R\sigma}} \left[f_{L\sigma} (\epsilon_0, V) - f_{R\sigma} (\epsilon_0, V) \right],
\end{eqnarray}
where $\displaystyle f_{X\sigma} (\epsilon_0, V) = \left[ 1+ e^{\beta(\epsilon_0 -\mu_{X\sigma} \pm eV/2)}\right]$ corresponds to the Fermi-Dirac distribution of the left (X=L) and right (X=R) electrodes with spin dependent chemical potential $\mu_{\sigma} =  \sigma \alpha_{A} \cos\theta$, $\alpha_{A}$ being the maximal effective Zeeman coupling in the Au electrode, and $\sigma = \pm 1$ for spin-up and -down, respectively (up and down are now defined in the $z-$direction). Here, $\gamma_L$ and $\gamma_{R\sigma}$ are the tunneling rates between the L and R-electrode and the molecular level, respectively, and $\alpha_{A}$ is the spin-orbit coupling (SOC) strength of the metallic electrode, and $\theta$ is the angle between the surface magnetization in the metallic electrode and the principle axis of the molecules. We simplify the expression of the current in \eqref{spin_dep_curr} as 
\begin{equation}
    I_{\text{low}(\text{high})} = \frac{\gamma}{2}  e^{-\beta \epsilon_{A}} e^ {\left[ \pm \beta \alpha_{A} \cos\theta \right]}
\end{equation}
for spin-down (low) and spin-up (high) electrons, respectively. In this simplification we have assumed that $\gamma_{L} = \gamma_{R\sigma} = \gamma$ and $\epsilon_{0} = \epsilon_{A}$, i.e., LUMO of the single level is equivalent (and equal) to the thermal activation. The total current and spin-polarization are given by $I = I_{\text{high}} + I_{\text{low}}$ and $I_S = I_{\text{high}} - I_{\text{low}}$, respectively.

Similarly, the densities of spin-up (-down) electrons are given by,
\begin{eqnarray}\label{charge_den}
n_{\sigma} &=& \frac{1}{2} \left[f( \epsilon_0 - \sigma\alpha_A \cos \theta + V/2 ) \right. + \\ \nonumber
& & \left. +f( \epsilon_0 -\sigma\alpha_A \cos \theta - V/2 )\right].
\end{eqnarray}
We linearize \eqref{charge_den} around $(V, \beta, \epsilon_{0}=\epsilon_A , \alpha_A ) = (0,0,0,0)$ and find the spin-polarization
$\Delta S = n_{\uparrow} -n_{\downarrow} \approx \frac{1}{2}\beta \alpha_{A} \cos\theta $. Once the expresions for the magnetization-dependent currents 
and spin-densities are obtained, one can proceed with the self-consistent calculation described in Sec.~\ref{sec:spinterface_model} (and in detail in Refs.~\cite{alwan2021spinterface,dubi2022spinterface}). We use the data from Q22 to fit the numerical parameters of the model, which are $\gamma=300$ nA, $\epsilon_A=12$ meV. The CISS fit parameters of the spinterface model are found to be $\alpha_A=0.0012$eV and $\alpha_1=0.98$eV. While these are very different from typical values found is e.g., molecular junctions \cite{dubi2022spinterface}, a possible reason is that the effective contact for the chiral molecules in Q22 is not Au (as it is in typical molecular transport measurements of the CISS effect) but either TaS$_2$ or TiS$_2$, which may have strikingly different interface properties than Au.

\begin{figure}
\includegraphics[scale=0.21]{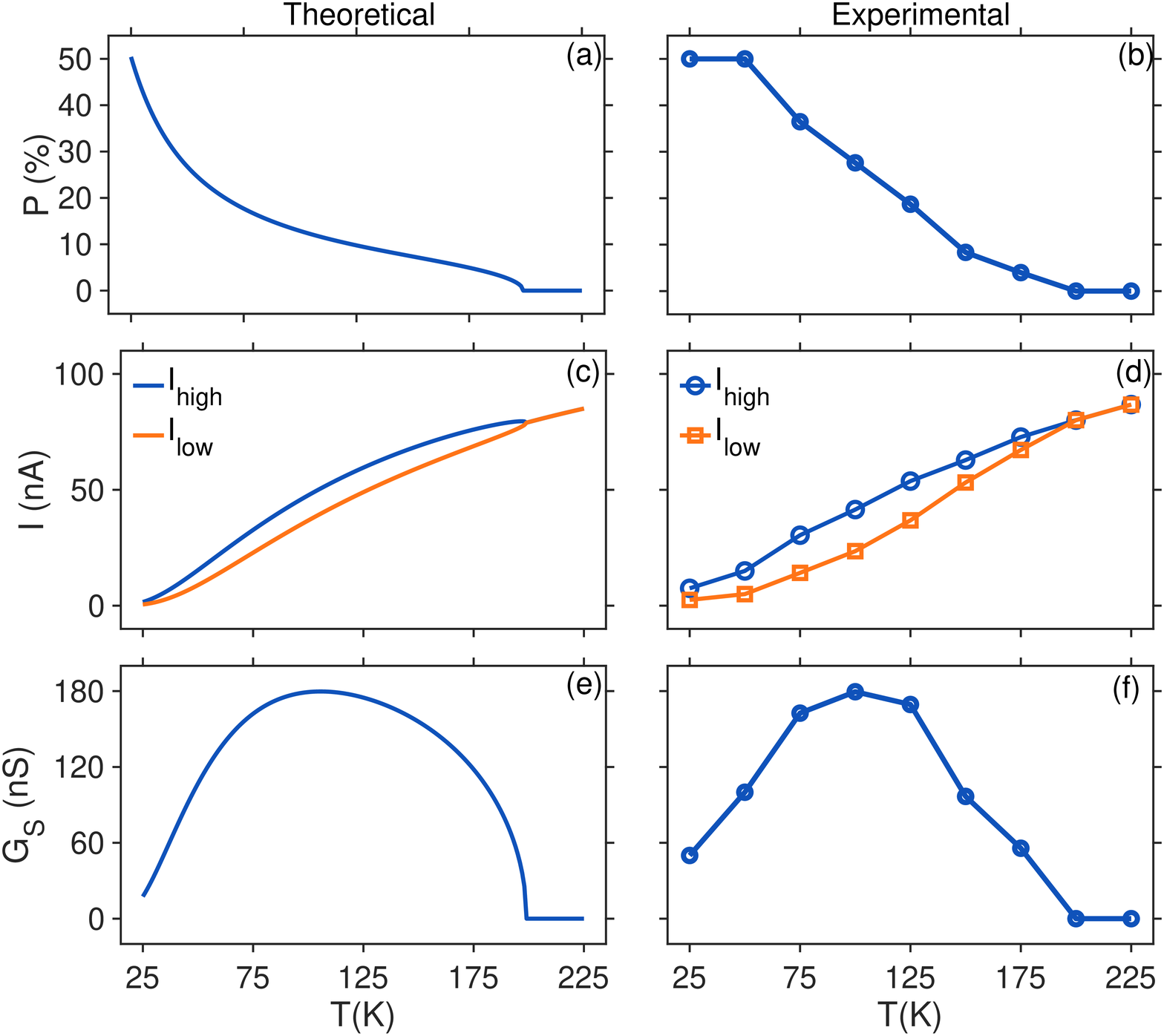}
 \caption{(a) Calculated polarization (in $\%$), (c) current (in nA) and (e) spin conductance (in nS) as functions of temperature ($K$), based on the theoretical approach of Ref.~\onlinecite{alwan2021spinterface}. The experimental data points, extracted from Qian, et.al., \cite{qian2022chiral}), (b) polarization (in $\%$), (d) current (in nA) and (f) spin conductance (in nS) as functions of temperature ($K$). The  qualitative resemblance between theory and data is clearly visible. 
} \label{fig:prediction_vs_expt}
\end{figure}

The results of the calculation are presented in Fig.~\ref{fig:prediction_vs_expt}, showing the polarization $P$  (Fig.~\ref{fig:prediction_vs_expt}(a)), the high and low currents (Fig.~\ref{fig:prediction_vs_expt}(c)) and the spin conductance $G_s$ (Fig.~\ref{fig:prediction_vs_expt}(e)), all as a function of temperature. The right-hand panels (b,d,f) show the experimental data. The qualitative resemblance between the theory and the data is clearly visible. 

\section{Summary and conclusions}
 In summary, we discussed the temperature-dependence of the CISS effect as a means to differentiate between different theoretical suggestions for the physical origin of the CISS effect. Specifically, one can point to two main theoretical ``branches". The first, based on electron-phonon interactions (and represented by, e.g., Ref.~\onlinecite{fransson2020vibrational}) suggests that the CISS effect increases with increasing temperatures and vanishes at low temperatures. The second, represented by, e.g., the spinterface model of Ref.~\onlinecite{alwan2021spinterface}, suggests that the CISS effect is finite at low temperatures and decreases with increasing temperatures. 

 Specifically, we showed how the CISS effect vanishes within the spinterface model, considering a single-level molecular junction within the full Landauer model, and discussed how the temperature-dependent CISS effect should be quantified. 

 Finally, we discussed in some detail the temperature-dependence of the CISS effect as it arises from the recent experiments of Qian, et.al., \cite{qian2022chiral}. These authors demonstrated that they are able to fabricate a remarkable, robust solid-state chiral material platform, based on chiral molecules intercalated between two-dimensional atomic crystals, which exhibits the CISS effect. 
 
 However, a consistent analysis of their data implies that, in opposite to their suggestion, the CISS effect is finite at low temperatures and is reduced with increasing temperature. Thus, their data is more consistent with the ``spinterface" model (which shows remarkable agreement to the experimental data) than with the phonon mechanism for the CISS effect. 
 
 We encourage the experimental community of the CISS effect to further explore the temperature dependence of the CISS effect, both in molecular transport setup and in photo-emission experiments \cite{mollers2022chirality}. Observing the CISS effect at low temperatures and its temperature-dependence in different systems (different molecules, different electrodes, etc.)  are necessary to help the community to fully understand the origin of the CISS effect.

 \section{Supplementary Material}
 Calculation details is available in the supplementary material.



\bibliography{main, refs}
\end{document}